\documentclass[11pt]{article}

\usepackage{amsmath,amssymb,bbm,amsfonts,amsthm}
\usepackage{commath}
\usepackage{graphicx}
\usepackage[document]{ragged2e}
\usepackage{pagecolor,color}

\newcommand{\noun}[1]{\textsc{#1}}

\usepackage[utf8]{inputenc}
\usepackage[T1]{fontenc}
\usepackage[margin=1.8cm]{geometry}

\usepackage{multicol}
\usepackage{setspace}

\usepackage{natbib}
\usepackage[french]{babel}

\def \draft {1}

\usepackage{xparse}
\usepackage{ifthen}
\DeclareDocumentCommand{\comment}{m o o o o}
{\ifthenelse{\draft=1}{
    \textcolor{red}{\textbf{C : }#1}
    \IfValueT{#2}{\textcolor{blue}{\textbf{A1 : }#2}}
    \IfValueT{#3}{\textcolor{ForestGreen}{\textbf{A2 : }#3}}
    \IfValueT{#4}{\textcolor{red!50!blue}{\textbf{A3 : }#4}}
    \IfValueT{#5}{\textcolor{Aquamarine}{\textbf{A4 : }#5}}
 }{}
}

\newcommand{\todo}[1]{
\ifthenelse{\draft=1}{\textcolor{red!50!blue}{\textbf{TODO : \textit{#1}}}}{}
}



\begin{document}

\title{Space and complexities of territorial systems
\\\bigskip
\textit{Actes des Journ{\'e}es de Rochebrune 2019}
}
\author{\noun{Juste Raimbault}$^{1,2,3}$
\bigskip\\
$^1$ UPS CNRS 3611 ISC-PIF\\
$^2$ CASA, UCL\\
$^3$ UMR CNRS 8504 G{\'e}ographie-cit{\'e}s
}
\date{}


\maketitle

\justify

\begin{abstract}
	The spatial character of territorial systems plays a crucial role in the emergence of their complexities. This contribution aims at illustrating to what extent different types of complexities can be exhibited in models of such systems. We develop from a theoretical viewpoint some arguments illustrating ontological complexity, in the sense of the diversity and multidimensionality of possible representations, and then complexity in the sense of emergence, i.e. the necessity of the existence of several autonomous levels. We then propose numerical experiments to explore properties of complexity (dynamical complexity and co-evolution) within two simple models of urban morphogenesis. We finally suggest other dimensions of complexity which could be typical of territorial systems.\medskip
\end{abstract}

\section{Introduction}

Territorial systems, which can be understood as self-organized socio-spatial structures  \citep{pumain1997pour}, are an archetypal of complex systems studied by numerous approaches which include more or less the spatial aspect of these systems. Our interpretation of territorial systems is more particularly situated within the heritage of the evolutionary urban theory~\citep{pumain2018evolutionary} which understands urban systems as multi-level systems, in which the co-evolution of multiple components and agents determine their dynamics~\citep{raimbault2018caracterisation}. The complexity of these systems is thus closely linked to their spatial nature, since their dynamics are carried within the spatial distribution of their entities and sub-systems, and interactions between agents causing the emerging behavior are embedded in space.

The concept of complexity corresponds to various approaches and definitions, which depend on the domains in which they were introduced. For example, \cite{chu2008criteria} reviews the conceptual and operational approaches linked to the domain of artificial life and shows that there is no unified concept. \cite{deffuant2015visions} also develops different visions of complexity, ranging from a complexity close to a computational complexity to an irreducible complexity typical of social systems. A previous work by \cite{raimbault2018relating} has proposed to suggest epistemological bridges between different approaches and definitions of complexity, and more particularly the links between complexity in the sense of emergence, computational complexity and informational complexity. As highlighted by \cite{batty2018defining}, dimensions of complexity are potentially infinite and an approach which would be universally valid for all problems and systems can not exist. It is therefore crucial to combine different approaches of complexity to understand its role within territorial systems.

This contribution aims at illustrating an approach of territorial systems through urban geography, and to what extent their complexity is closely linked to their spatial nature. We propose here to illustrate the links between complexity and space according to different views of it, both through theoretical considerations but also through the exploration of simulation models of territorial systems. The rest of this paper is structured as follows: first in two sections we develop a conceptual approach to ontological complexity and emergence in territorial systems. We then describe results from the simulation of an urban morphogenesis model which exhibits properties typical of dynamical complexity. An other experiment allows then to show the link between complexity in the sense of a co-evolution and the spatial structure of the system. We finally discuss the implications of these results and possible developments.

\section{Ontological complexity}

A first theoretical approach allows to tackle what we call \emph{ontological complexity}, which was proposed by \cite{pumain2003approche} as the broadness of disciplinary viewpoints necessary to apprehend a common object. The multidimensionality of territorial systems remains a main issue for their understanding, as show \cite{perez2016agent} in the case of multi-agent systems. Thus, interdisciplinarity would be intrinsic to ``urban sciences'', i.e. to the different disciplines considering the city as an object of study, although it is in practice difficult and subject to the contingency of the social organisation of sciences \citep{dupuy2015sciences}.

We take the example of \cite{raimbault2017invisible} as a proof-of-concept of the diversity of approaches possible in the case of interactions between networks and territories, and suggest hypotheses regarding the role of space in this complexity, such as the evolutive processes of diversification and specialization linked to spatial niches. More precisely, \cite{raimbault2017invisible} establishes a scientific map of modeling approaches to interaction between transportation networks and territories. Through a citation network analysis, complementary approaches are exhibited, including for example quantitative geography, political geography, urban economics, planning, physics. Each approach provides a particular dimension of the same system. Space leading to the differentiation of sub-systems, it seems on the one hand natural that some have a higher importance than others in the different instances, and on the other hand that complementary questionings coexist (these being in fact endogenous to territorial systems, since innovation and thus research and science are an important driver of urban systems dynamics \citep{pumain2010theorie}).

At this point we propose an hypothesis, which empirical exploration would require deeper scientometric analysis out of the scope of this work, according to which a more elaborated spatialization would be linked to a broader ontological spectrum. The example of \emph{New Economic Geography} and geographical economics illustrates that aspect \citep{marchionni2004geographical}: the first approach, in order to deploy its analytical tools, imposes a strong reductionism on space (line or circle for a large part of models) and on objects (representative agents, homogeneity), whereas the second will favor empirical descriptions more faithful to geographical particularities. It is difficult to know if space is richer because the approach is not reductionist or the contrary, that a rich space increases ontological range. Pretend a sense of causality would in fact be counter-productive, and this congruence confirms our argument that an elaborated spatialization of models of territorial systems is correlated to a higher ontological richness.

This reflexion can be shifted to the methodological domain, within which we can highlight the tension between model precision and robustness of results, in particular by comparing agent-based models and dynamical systems models allowing a certain level of analytical tractability. The archetype of approaches in economic geography, considering an unidimensional system \citep{krugman1992dynamic}, looses the essence of space which is heterogeneity and multiplicity of alternatives, whereas precisely spatialized models such as Luti models, will necessitate simulations and a restricted validation of results \citep{bonnel2014survey}. The degree of precision of models is however not directly linked to their spatial character, as shows the comparison between agent-based models and demographic microsimulations \citep{birkin2011spatial}. However, approaches including an elaborated description of space, with the inclusion of scales for example, are mainly associated to an ontological richness in the study of these systems, which need to be maintained to ensure the diversity of knowledges on the subject \citep{sanders2018survival}.

\section{Complexity and emergence}

Our second theoretical entry focuses on complexity as the weak emergence of structures and autonomy of upper levels~\citep{bedau2002downward}. Weak emergence consists in the interpretation that the levels of a systems are effectively \emph{computed} \citep{morin1980methode} by itself, and that it is necessary to simulate the system to reproduce them. The diachronic aspect of dynamics allows this view of emergence to not being incompatible with downward causation. We recall the intrinsic multi-scalar aspect of territorial systems, which is highlighted for example in the approach of cities as systems within systems of cities, as introduced by \cite{pumain1997pour} which extends \cite{berry1964cities}. Furthermore, there is currently a necessity to produce spatial models integrating effectively this multi-scalar aspect~\citep{rozenblat2018conclusion}, with the aim of effectively operational models.

The difficulty to endogenize autonomous upper levels can for example be illustrated by~\cite{lenechet:halshs-01272236} which introduces a co-evolution model between transportation and land-use at the metropolitan scale integrating an endogenous governance structure for the transportation network. Simulating tractations between local actors of transportation, some regimes lead to the emergence of an intermediary governance level produced by the collaboration between neighbor actors. The three decision levels are then indeed ontologically autonomous but also in terms of their dynamics. The emergence of collaboration is finely linked to the spatial structure, since actors include accessibility patterns in their decision-making process. Less rich models of co-evolution between transport and land-use, as \cite{raimbault2018urban} at the mesoscopic scale taking into account urban form and topology of the road network, or \cite{raimbault2018modeling} which abstract networks as distance matrices at the macroscopic scale and postulates dynamics on these, indeed capture an emergence fundamentally linked to space, but remain restricted in the articulation of levels, whereas the governance model is closer to the qualitative emergence of upper levels.

Systems with a lower ontological complexity also exhibit a crucial role of weak emergence. The local state of a traffic flow is partly a consequence of the global state of the system, in particular when important congestion patterns can be observed at the macroscopic scale. In that case, spatio-temporal patterns are again central to the emergence process \citep{treiber2010three}. In these different examples, space again play a crucial role for the presence of complexity, since auto-organization implies a spatial distribution of agents, and the levels of emergence are directly linked to spatial scales whatever the interpretation of these \citep{manson2008does}. An approach of auto-organization through morphogenesis, that will be detailed within models in the following, furthermore insists on the spatial aspect in emergence, since it links form and function \citep{doursat2012morphogenetic}, the first being necessarily spatialized.

\section{Dynamical complexity}

In this section and the following, we propose to use simulation models of urban morphogenesis to show in a concrete manner other complexities of territorial systems. The models, detailed in the following, are implemented in Netlogo \citep{wilensky1999netlogo} and in scala, and explored using the model exploration software OpenMOLE \citep{reuillon2013openmole}. All code and results is openly available on the git repository of the project at \texttt{https://github.com/JusteRaimbault/SpatialComplexity}. Simulation results are available at \texttt{https://doi.org/10.7910/DVN/LENFVH}.

The understanding of complex systems as dynamical systems with more or less chaotic attractors has been largely developed in geography \citep{dauphine1995chaos}. For example, \cite{e18060197} considers that the semantic information of an urban environment is linked to the attractors of dynamical systems ruling the dynamics of its agents. This type of approach can moreover be linked to fractal approaches of urban systems, initially introduced by \cite{batty1994fractal}, and for example more recently applied to real urban planning problems \citep{yamu2015spatial}. These questions are more generally linked to transversal questions on spatio-temporal chaos \citep{crutchfield1987phenomenology}. The understanding of the link between time and space, and the corresponding non-uniform, non-stationary, non-ergodic dynamics, remains in construction on the theoretical, methodological and empirical dimensions, and promises numerous applications for the study of territorial systems. For example, \cite{chen2009urban} combines cross-correlations and Fourier analysis to study an urban gravity model. An important research direction in that frame is the understanding of the non-stationarity of territorial systems properties, and \cite{raimbault2018urban} explores it in the case of urban morphology and network topology.


We use here an urban morphogenesis model introduced by \cite{raimbault2018calibration} to illustrate the properties of sensitivity to initial conditions and path-dependancy of territorial systems~\citep{pumain2012urban}. In particular, we show the strong sensitivity of final simulated urban forms to spatial perturbations, and more generally the path-dependancy of trajectories for aggregated morphological indicators. This relatively simple model simulates population growth on a grid, through the iterative addition of population, which localizes following a preferential attachment to the already distributed population and diffuses in space. Important parameters are $N_G$ the exogenous growth rate, $P_{max}$ the total final population, $\alpha$ the preferential attachment exponent, $\beta$ the diffusion rate and $n_d$ a number of diffusions for each time step.

The experiment done by \cite{raimbault2018calibration} on a simplified one-dimensional version of the model shows that semi-stationary distributions of populations can be at a maximal distance (in a sense of a L2 norm between distributions) starting from a same initial configuration. In two dimensions, the phenomenon is identical as shown by the experiment done here. The model is simulated starting from an initial configuration which is particularly sensitive, constituted by four initial centers of similar size. For a grid of size 100, four centers are positioned in the middle of each cadrant, with an exponential kernel for population of the form $P_0 \cdot \exp \left(-r/r_0\right)$ with $P_0 = 100$ and $r_0 = 5$. The model is then simulated for given values of parameters for aggregation $\alpha$, diffusion $\beta, n_d$, growth and population $N_G, P_{max}$ (voir \cite{raimbault2018calibration}), and a realization of the distance between configurations $d(t)$ is computed on two independent realizations of populations $P^{(k)}_i(t)$, as $d(t)=\norm{P^{(1)}_i(t) - P^{(2)}_i(t)}$. A direct experience plan is achieved, with a LHS sampling of 390 parameter points and 1000 repetitions.

\begin{figure}
	\includegraphics[width=0.49\linewidth]{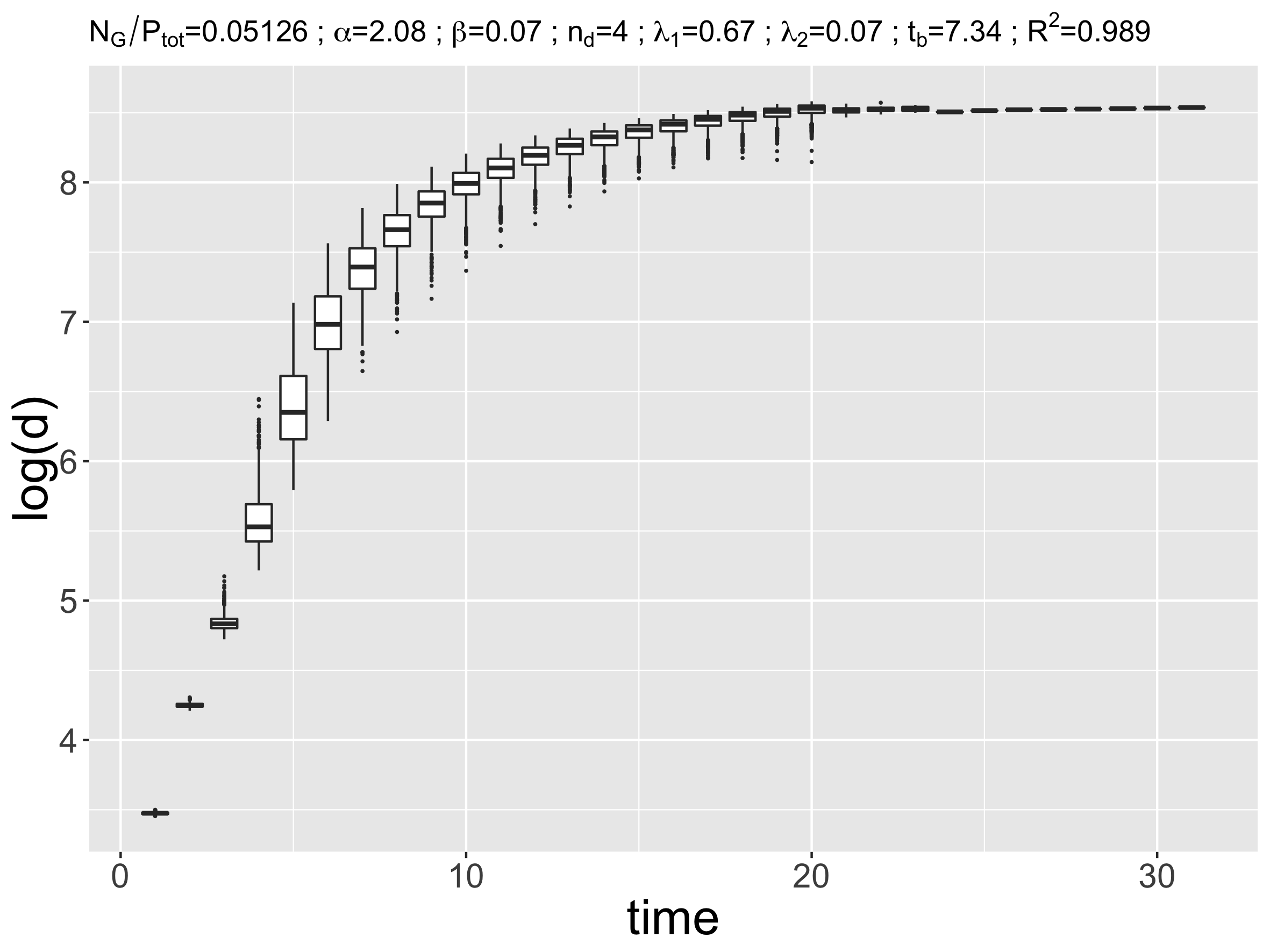}
	\includegraphics[width=0.49\linewidth]{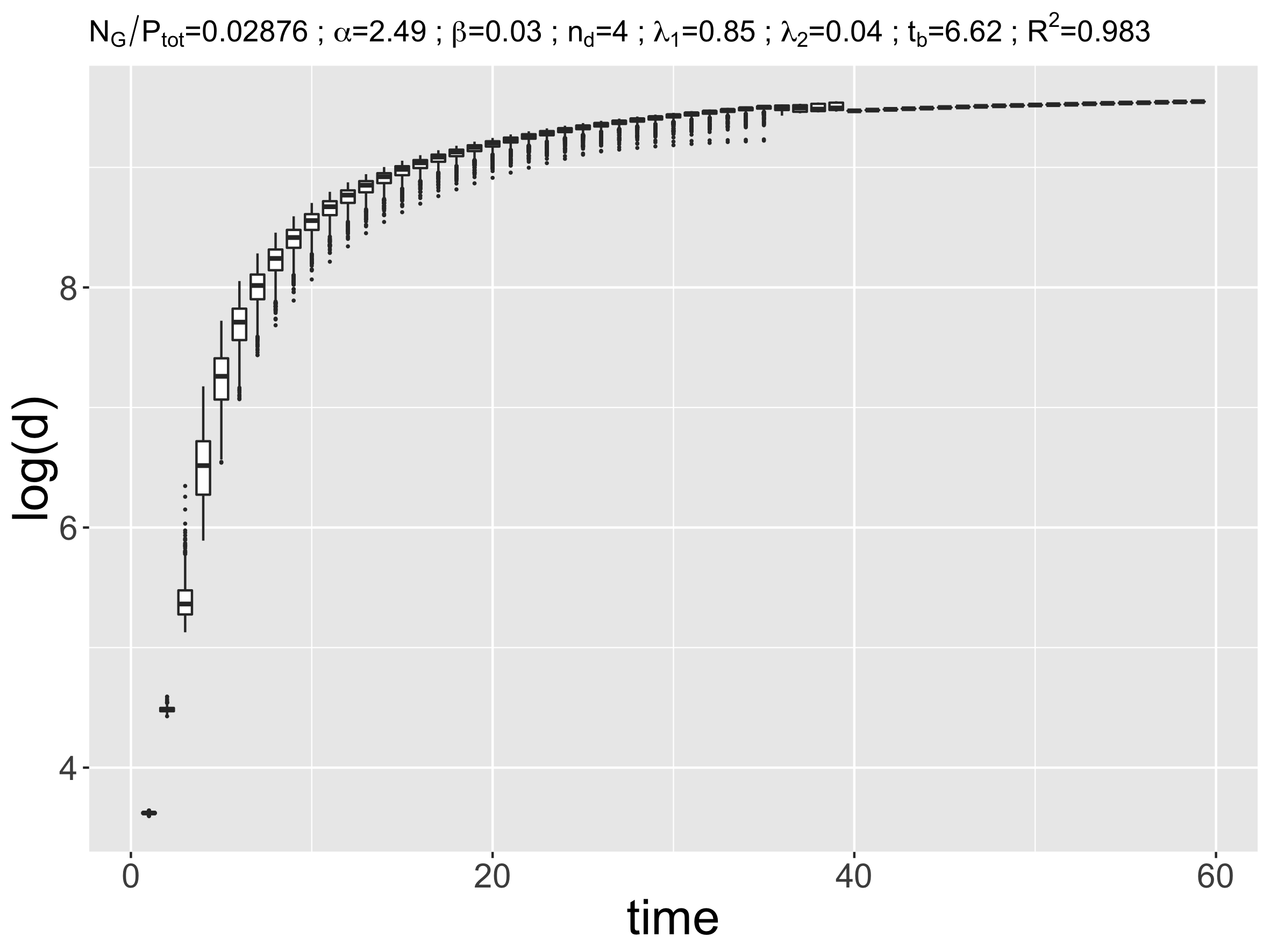}
	\includegraphics[width=0.49\linewidth]{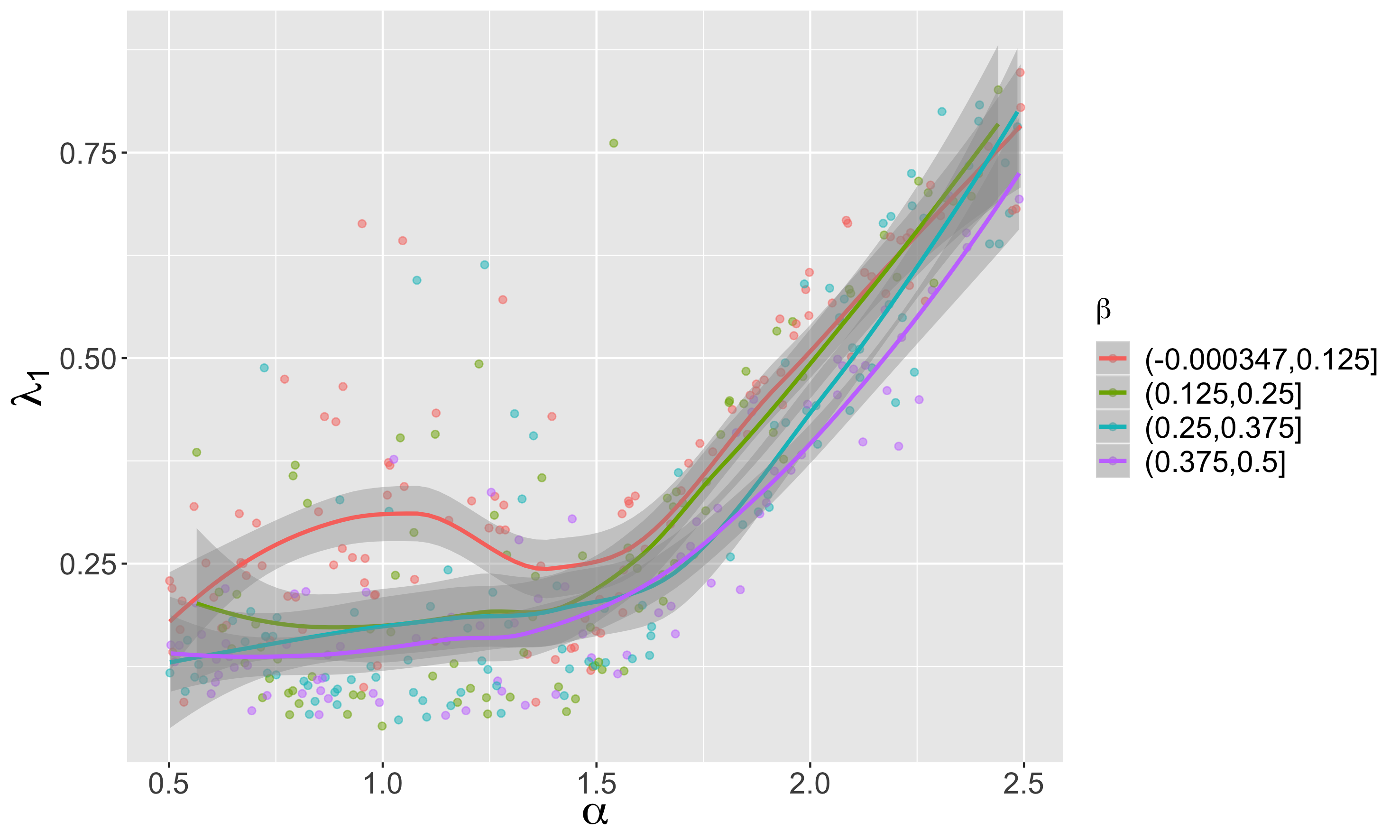}
	\includegraphics[width=0.49\linewidth]{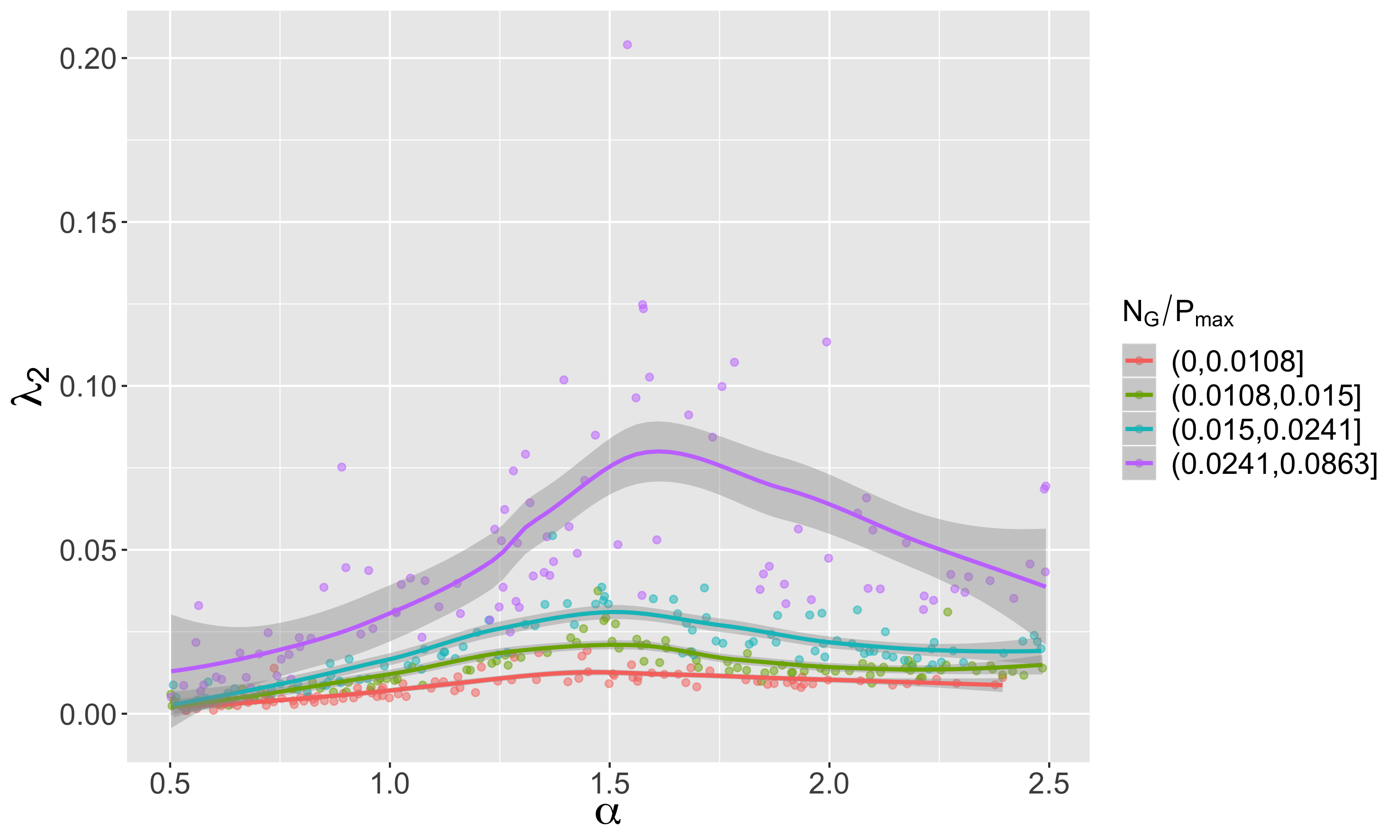}
	\caption{\textbf{Estimation of local Lyapounov exponents for the reaction-diffusion model.} \textit{(Top)} Temporal profiles as descriptive statistics for $\log d(t)$ as a function of $t$, for two configurations showing a maximal fit and a maximal $\lambda_1$ value ; \textit{(Bottom left)} Value of $\lambda_1$ as a function of $\alpha$, for different values of $\beta$ (color) ; \textit{(Bottom right)} Value of $\lambda_2$ as a function of $\alpha$, for different values of $N_G / P_{max}$.}
	\label{fig:lyapounov}
\end{figure}

We show in Fig.~\ref{fig:lyapounov} a basic estimation of Lyapounov exponents, achieved with piecewise linear regression estimation on two segments, of $\log d(t)$ as a function of $t$, on all repetitions for one parameter point. The adjustment is relatively good since the first quartile of the adjusted $R^2$ is at 0.92 and its average at 0.93. The Lyapounov exponent on the first segment, that we write $\lambda_1$, varies between 0.05 and 0.84 with an average of 0.31. Thus, the quality of the adjustment and the positive values show the strong sensitivity to initial conditions of generated configurations. Temporal regimes are quite different, since the estimated breakpoint time $t_b$ varies between 2.5 and 37.0. Values of $\lambda_2$ are lower, with an average at 0.02. There exists a first regime during which the most of the urban structure is constituted followed by a second regime (that we can describe as ``semi-stationary'') during which evolutions are smaller. This moreover suggests an endogenous stopping criteria for the model. The curves of $\log d(t)$ shown in Fig.~\ref{fig:lyapounov} are for a maximal value of the adjustment and a maximal value of $\lambda_1$. We also show the variations of exponents as a function of parameters. $\lambda_1$ as a function of the aggregation $\alpha$ exhibits an unexpected behavior, relatively stable below 1.5 and then a linear increase above, without modification around 1, what suggests a threshold over which this process dramatically influences urban form. For the lower values of $\beta$, a local maximum at 1 suggests a role of the infra or supra-linear aspect of the regime. For the behavior of $\lambda_2$, a significant effect is observed only for the larger growth rates, and corresponds to a maximum around 1.5, i.e. that the switch to the second regime for $\lambda_1$ corresponds in this case to a better stabilization for the second phase. This study shows thus the sensitivity of urban configurations to infinitesimal perturbations (what is equivalent to a perturbation of the initial state).

\begin{figure}
	\centering
	\includegraphics[width=0.7\linewidth]{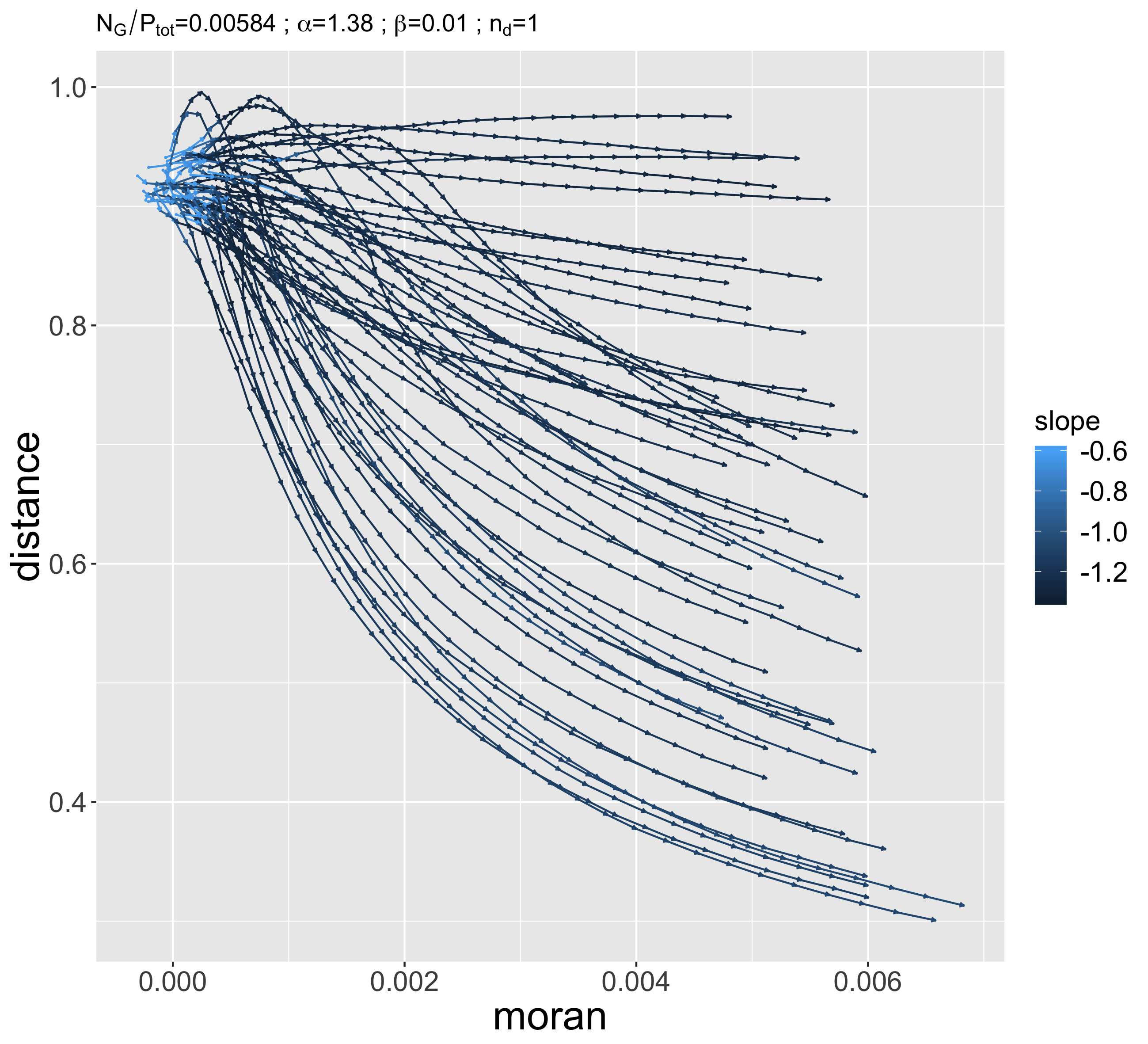}
	\caption{\textbf{Temporal trajectories of the morphological indicators.} We show for a parameter point a sample of 50 trajectories in the morphological space (moran,distance), the color giving the level of hierarchy. In this configuration, the final variance is very large and trajectories sometimes cross others, confirming the path-dependence for morphological indicators.}
	\label{fig:morphotraj}
\end{figure}

Furthermore, we illustrate the temporal trajectories of aggregated morphological indicators. A similar experience plan with a single model simulation at each repetition and an empty initial state, but following in time the values of urban form indicators (Moran, entropy, average distance, hierarchy), shows that their trajectories are path-dependent. Indeed, the example of Fig.~\ref{fig:morphotraj} shows a significant divergence of trajectories starting from the same initial point, but more importantly some trajectories crossing (indicators may therefore not be formulated as a deterministic dynamical system in which trajectories may not cross because of the Cauchy-Lipschitz theorem), exhibiting numerous points where urban form is very close (at least for the two indicators shown) for two trajectories, but the past and the future are different. Therefore, the future state of an urban system entirely depends on the past trajectory, and two systems which are very similar at a given time may exhibit fundamentally different trajectories. It is naturally an illustration with a toy model, on the particular aspect of urban form, but the path-dependence effects are a priori even more important when taking into account the multiple levels, political aspects, and infrastructures \citep{pumain2012urban}.

\section{Complexity and co-evolution}

A certain approach of complex territorial systems focuses on the concept of co-evolution. The strong entanglement of elements in presence of what can be understood as territorial niches, in the sense of ecological niches of \cite{holland2012signals}, is an expression of a co-evolution and thus a complexity within these niches. \cite{raimbault2018modeling} shows the emergence of these spatial niches in a co-evolution model between cities and transportation networks at the macroscopic scale.

We thus explore here through simulation experiments the link between spatial non-stationarity, which is also a clue of spatial complexity, and emergence of niches in an hybrid morphogenesis model coupling urban and network development, introduced by~\cite{raimbault2014hybrid}.

The RBD model~\citep{raimbault2014hybrid} couples in a simple way urban growth and evolution of the road network. The flexibility of regimes it allows to capture provides in~\cite{raimbault2017identification} a test for a method to identify spatio-temporal causalities. We extend here this method through an endogenous detection of spatial areas corresponding to the different regimes, in order to show the emergence of niches through non-stationarity. For a precise description of the model together with its use to produce synthetic data, refer to \cite{raimbault2018caracterisation}. In our case, variable parameters are the initial number of centers $N_C$ and also the relative weights of the different explicative variables $(w_c,w_r,w_d)$ (distance to the center, distance to the network, density) determining the local value during the urban sprawl.


Non-stationarity is introduced by making these weight parameters vary in space. We distinguish two implementations, given values attributed to each center: (i) the local value of weights is given by the value for the closest center; (ii) the local value is the average of center values weighted by the distance to these.

Spatial niches are detected using an unsupervised classification on lagged correlation profiles estimated locally in space, i.e. $(\rho_{\tau}\left[X_i,X_j\right])_{\tau,i,j}$ where $- \tau_M \leq \tau \leq  \tau_M$ with $\tau_M = 5$, for the three couples of variables such that $i < j$. Time series are truncated below $t_0 = 5$ and at a given spatial point, correlations are estimated on cells in a radius of  $r_0 = 10$, with a spatial step of $\delta x = 5$ between estimation points. A k-means algorithm is used to classify profiles, with a number of clusters $k = N_C$. To remove the stochasticity of the classification, it is repeated $b = 1000$ times, and performance measures are estimated on all these realizations of the classification.

In order to quantify the classification, a solution could be to study a distance to the partition defined by stationarity areas. However, the computation of a distance between partitions is a NP-hard problem \citep{day1981complexity} for which even the optimal solutions \citep{porumbel2011efficient} are too computationally costly given the number of realizations. We use therefore the following indicators, capturing expected properties of spatial niches: (i) cumulated distance between centroids of the classification and centers, corrected by the distance between centroids and the distance between centers; (ii) average normalized radius of clusters; (iii) average distance between feature vectors used for classification; (iv) average intra-cluster variance. Each indicator is computed on the classification obtained, and also on a null model obtained with a random redistribution of cluster labels between the points.

\begin{figure}
	\includegraphics[width=0.49\textwidth]{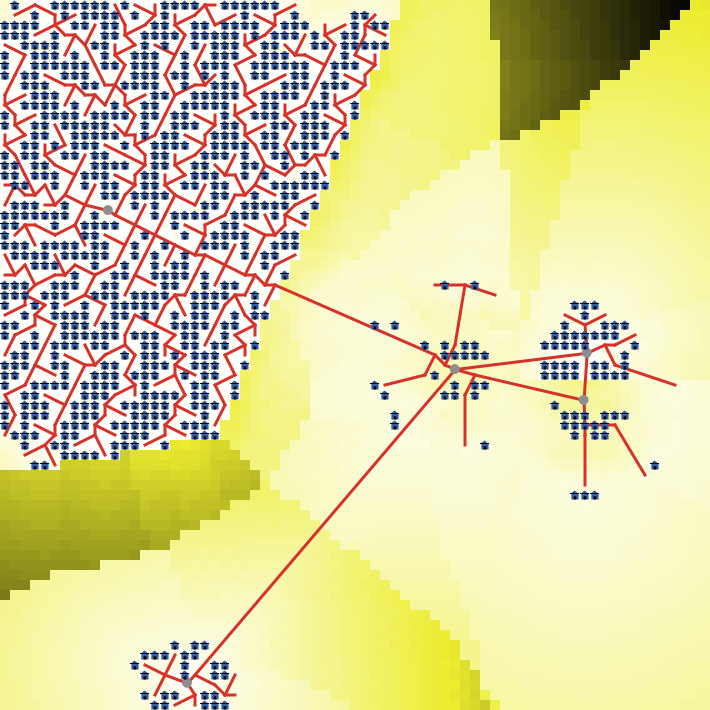}
	\includegraphics[width=0.49\textwidth]{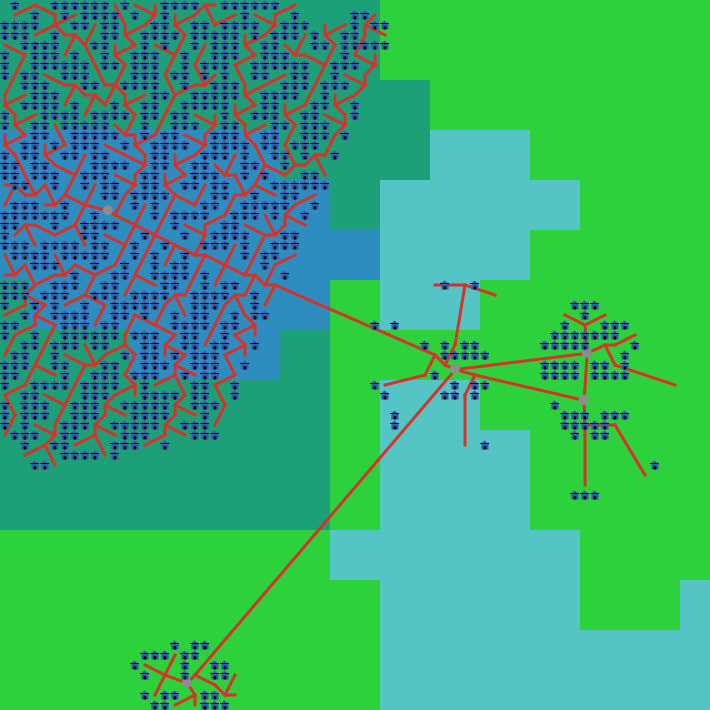}\\\bigskip
	\includegraphics[width=0.49\textwidth]{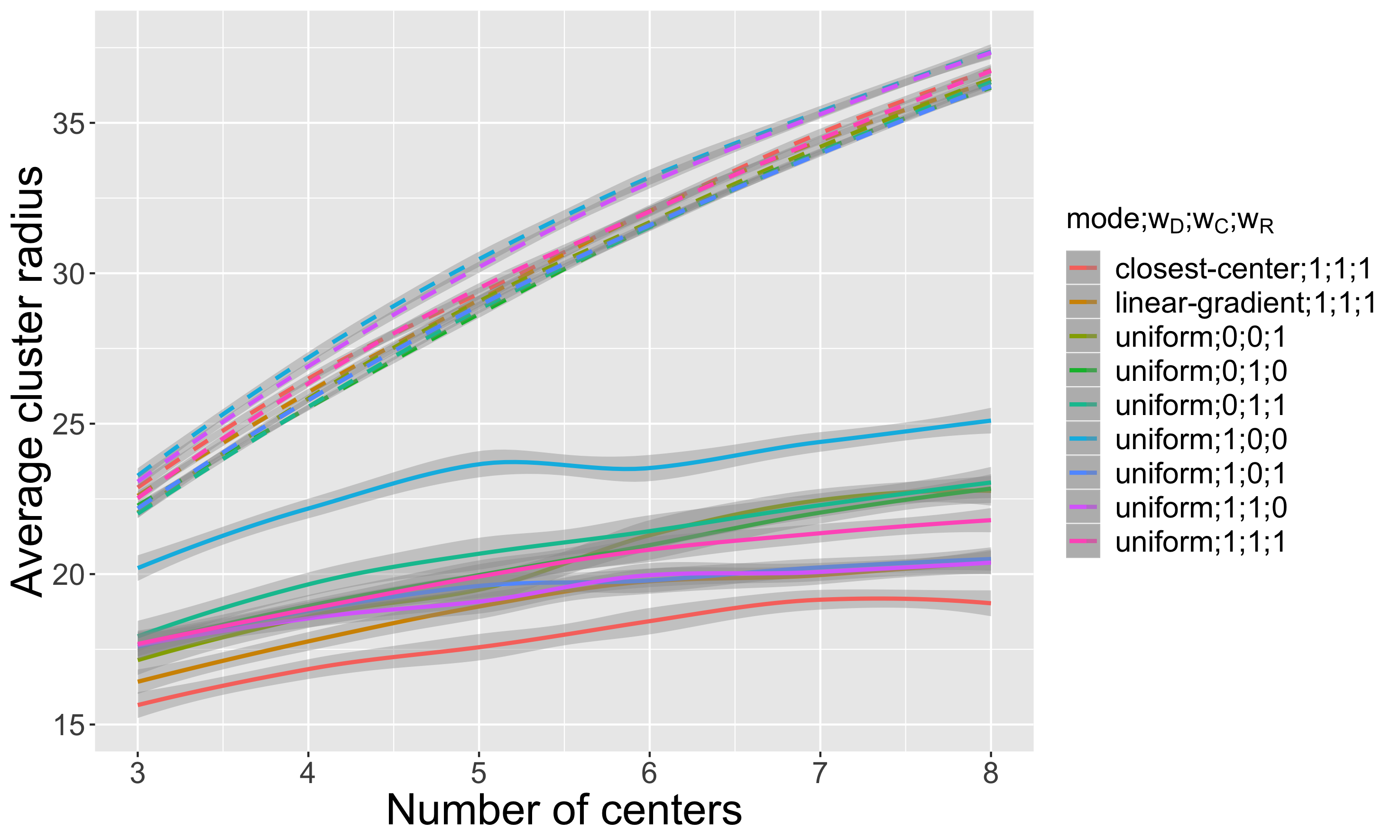}
	\includegraphics[width=0.49\textwidth]{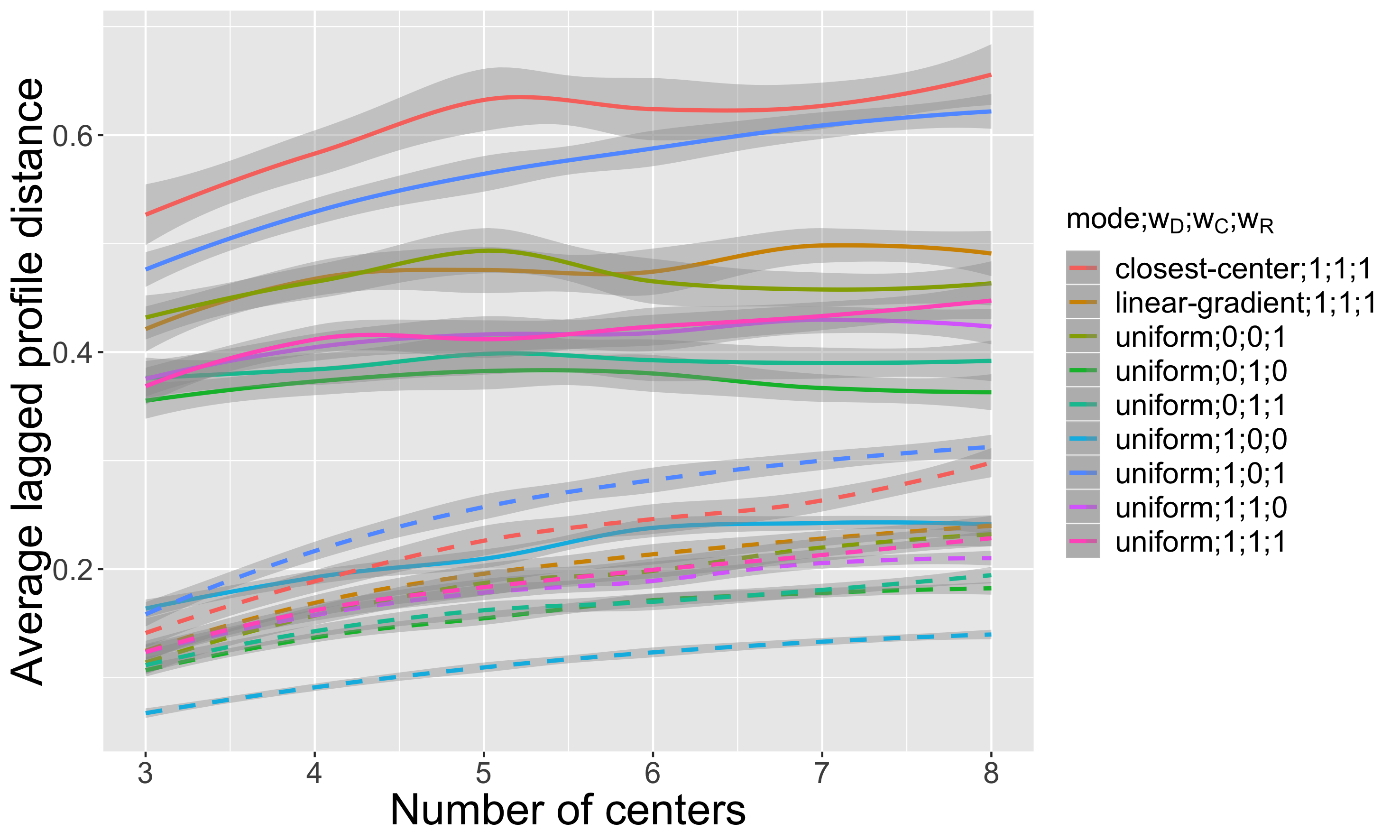}\\
	\includegraphics[width=0.49\textwidth]{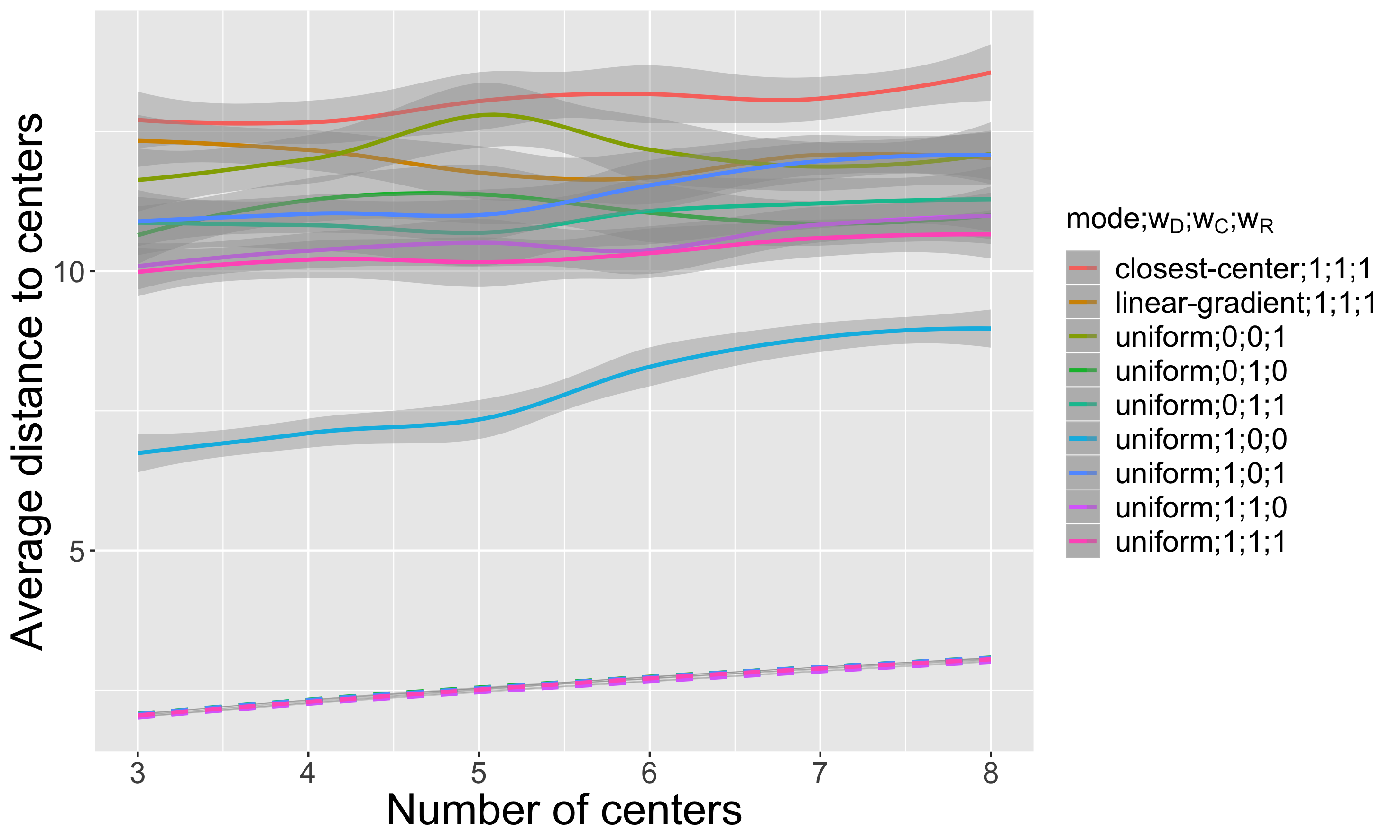}
	\includegraphics[width=0.49\textwidth]{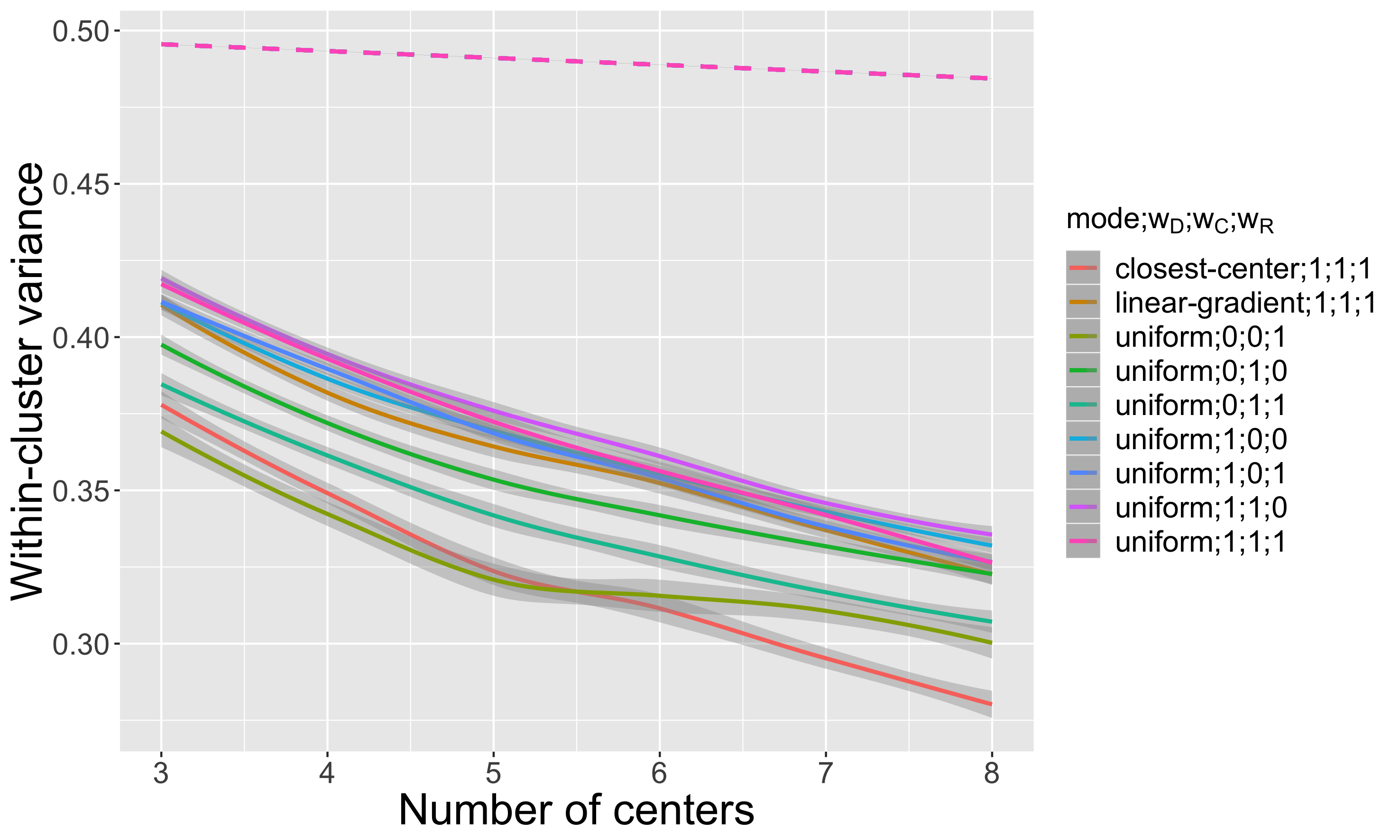}
	\caption{\textbf{Spatial niches of co-evolution in the RBD model.}\textit{(Top left)} Configuration generated with $N_C = 5$ centers and non-stationary parameters with closest centers, such that, in a decreasing vertical coordinate order, the weights for each center are $(w_d,w_r,w_c) = (0,1,0) ; (0,0,1) ; (1,0,1) ; (1,0,1) ; (0,0,1)$; \textit{(Top right)} Example of corresponding realization for the k-means classification, the color of cells giving the cluster; \textit{(Middle left)} Average normalized radius of clusters, for each mode and parameter configuration (color), and also for the null model of random clusters (dashed lines); \textit{(Middle right)} Average distance between lagged correlation profiles for cluster centroids; \textit{(Bottom left)} Average distance of centroids to centers; \textit{(Bottom right)} Intra-cluster variance.}
	\label{fig:rbd}
\end{figure}

Our experience plan compares the two modalities of non-stationarity (random distributions of weights for centers among the possible combinations of extreme values) to the uniform version of the model taken for all extreme values of weights, the number of centers $N_C$ varying between 3 and 8, and 100 replications are achieved for each parameter point. We show in Fig.~\ref{fig:rbd} an example of configuration obtained in the case of a closest center non-stationarity and the corresponding spatial classification. The spatial continuity of clusters is an important result since these correspond then to co-evolution niches. The values of indicators confirm the systematic existence of these niches. Indeed, the average radius, which gives the spatial sprawl of clusters, is firstly considerably different from the null model for all parameters, but moreover significantly (in the statistical sense) lower for the closest center non-stationarity compared to all control configurations. Regarding the linear non-stationarity, it appears to be not conclusive. Other indicators conform the emergence of niches: the distance between centroid profiles is the highest in the closest center non-stationarity case, what means that it is the case where clusters make most sense in terms of differentiating features. This also corresponds to the lowest intra-cluster variance, observed for the high numbers of clusters. The average distance of centroids to centers is however the highest in this configuration, showing that niches do not coincide with non-stationarity areas, what confirms that these are indeed emergent and not only the reproduction of the initial configuration. Thus, this experiment shows that spatial non-stationarity leads to the emergence of spatial niches (which have an intrinsic spatial consistence) for the lagged correlation profiles, and therefore of co-evolution niches in the sense of \cite{raimbault2018caracterisation}. Space is again crucial for the production of a complexity, this time in the sense of a co-evolution.

\section{Discussion}

Different approaches of complexity that we could not tackle can be suggested in perspective, as being also typical of territorial systems and with a potential link to space. There could for example exist a link between computational complexity and the nature of large deviation of territorial configurations (empirical impossibility in probability to obtain these with brute force): to what extent a territorial system is easy to generate with computation, and what properties can explain this possibility ? Indeed, similarly to computational properties of spatial self-organized biological systems such as the \emph{slime mould} which are able to solve NP-complete problems \citep{zhu2013amoeba}, territorial systems may in that sense have a certain intelligence, the question of how it is linked to its spatial configuration remaining open. A link between informational complexity and the diffusion of innovation in territorial systems can also be suggested \citep{favaro2011gibrat}. The diffusion of innovation is suggested as a fundamental driver of urban dynamics by the evolutive urban theory. Innovation is linked to cultural evolution processes \citep{Mesoudi25072017} and thus to an informational complexity in the sense of non-trivial patterns of transmission and processing of information. The spatial distribution of these processes, and to what extent it influences its properties, is an important research direction for the study of the spatial complexity of territorial systems.

The work developed here has been restricted to theoretical considerations and toy examples. The systematic construction, as much in the sense of constructing concepts through a systematic review, than the construction of theories and methods articulating in a relevant way these different links between complexity and space, remains an open perspective for future work. We thus conclude, following \cite{raimbault2017complexity}, that territorial systems are necessarily at the intersection of multiple complexities, and add, according to the different examples developed here, that their spatial nature plays an important role in their emergence.



\section*{Acknowledgements}

Results obtained in this paper have been computed on the virtual organisation \textit{vo.complex-system.eu} of the \textit{European Grid Infrastructure} (\texttt{http://www.egi.eu}). We thank the \textit{European Grid Infrastructure} and its \textit{National Grid Initiatives} (\textit{France-Grilles} in particular) to give the technical support and the infrastructure.





\end{document}